\documentclass[a4paper, 12pt]{article}
\usepackage{amsfonts,epsfig}
\usepackage{amsmath}

\newcommand{\re}{\textrm{e}}
\newcommand{\ri}{\textrm{i}}
\newcommand{\const}{\textrm{const}}

\begin{document}
\title{Surface  acoustic  waves in rotating orthorhombic crystals.}
\author{Michel Destrade}
\date{2004}
\label{firstpage}
\maketitle
\bigskip

%
\begin{abstract}
The propagation of surface (Rayleigh) waves over a rotating 
orthorhombic crystal is studied.
The crystal possesses three crystallographic axes, normal to the 
symmetry planes:
the half-space is cut along a plane normal to one of these axes,
the wave travels in the direction of another,
and the rotation occurs at a uniform rate about any of the three axes.
The secular equation for the surface wave speed is found explicitly;
in contrast to the non-rotating case, it is dispersive 
(frequency-dependent).
Both Coriolis and centrifugal accelerations appear in the equations of 
motion: none can be neglected in favor of the other, even at small
rotation rates.
\end{abstract}

\newpage

%

\section{Introduction}
The main applications of Surface Acoustic Wave (SAW) devices are in 
the field of wireless communication, where SAW components are used 
to generate and collect high frequency signals.
In mobile phones, global positioning systems, or color television sets,
these components are of course motionless.
There exist however situations where SAWs propagate over a rotating 
surface.
For instance, SAW sensors were used to monitor the pressure of a tire 
in a moving automobile (Pohl \textit{et al.} 1997);
an ultrasonic SAW-based apparatus was designed to detect flaws on the
surface of a rotating member such as a roller bearing 
(Kawasaki \textit{et al.} 1991);
robust and shock-resistant SAW resonator gyroscopes have been
miniaturized to a 1 cm $\times$ 1 cm surface 
(Jose \textit{et al.} 2000); etc. 
For the theoretical aspect, Schoenberg \& Censor (1973) seem to have 
been the first authors to describe Rayleigh waves in a rotating, 
homogeneous, isotropic, linear, elastic medium, for any orientation of 
the rotation axis with respect to the free surface, and with both the 
Coriolis and the centrifugal accelerations taken into account. 
They pointed out that, in contrast to the non-rotating case, bulk and 
surface waves are dispersive, and that the acoustical tensor is 
Hermitian instead of real symmetric; 
they did not however derive the secular equation for surface waves. 
Clarke \& Burdess (1994) obtained this equation for an isotropic 
half-space rotating about an axis orthogonal to the direction of 
propagation and to the normal to the half-space. 
Others (Lao 1980; Wauer 1999; Grigor'evski\u{i} \textit{et al.} 2000; 
etc.) considered similar problems but neglected the centrifugal 
acceleration. 
For anisotropic crystals, Fang \textit{et al.} (2000) and 
Zhou \& Jiang (2001) considered crystals of tetragonal symmetry but 
did not derive the secular equation explicitly. 
As pointed out in the latter article, the influence of the centrifugal 
acceleration upon the SAW speed is of the same order as that of the 
Coriolis acceleration and should not be overlooked, even at small 
rotation rates. 
In this paper, an analytical investigation of the propagation of SAWs 
in a rotating  orthorhombic crystal is presented. 

In \S2, the crystal considered is at first of general (triclinic) 
anisotropy and rotates at a uniform rate about any axis. 
The equations of motion and the boundary conditions are derived. 
Then the analysis is specialized to crystals with rhombic 
and higher symmetries. 
The material is cut along the plane of symmetry $x_2=0$ and the wave 
travels in the $Ox_1$ direction, where the $Ox_i$ axes are 
crystallographic axes. 
The semi-infinite body rotates along $Ox_1$, $Ox_2$, or $Ox_3$. 
In the latter case, anti-plane motions and stresses decouple from 
their in-plane counterparts; in the other two cases, they do not. 
The choice of orthorhombic symmetry is of great importance because, 
as pointed out by Royer \& Dieulesaint (1984), it covers 16 types 
of symmetries, including tetragonal, hexagonal, cubic, and of course 
isotropic; 
the choice of a rotation axis aligned with the normal to the free 
plane or to crystallographic axes in this plane is justified from an 
experimental point of view. 
In \S3, the secular (dispersion) equations are derived explicitly for 
each rotation, using the method of the polarization vector. 
This method was used by Currie (1979) and by Taziev (1989) in order 
to derive explicit secular equations for SAWs in non-rotating crystals.
 
Recently, Ting (2003$ab$) improved upon this method and placed it 
within the framework of the Stroh (1958, 1962) formalism. 
Here, the fundamental equations of the method are obtained in a manner 
which does not rely upon this formalism. 
The secular equation is obtained as a quadratic relationship between 
three determinants (of $2 \times 2$ matrices for rotation about 
$Ox_3$, of $3 \times 3$ matrices for rotations about $Ox_1$ or 
$Ox_2$), whose elements are given explicitly in terms of the elastic 
parameters, the wave speed, and $\delta$ --- the ratio of the rotation 
rate by the wave frequency. 
Several secular equations arise in the resolution of the problem 
but only one is kept; 
the others are dismissed on the argument that they do not reduce to 
the known single cubic secular equation when the rotation rate 
vanishes. 
Numerically, the Rayleigh wave speed decreases monotonically as 
$\delta$ increases. 
This influence is shown graphically for $\alpha$-Iodic Acid 
(HIO$_3$, rhombic). 
For this graph, $\delta$ is  chosen to vary from 0 to 10. 
Although it is not realistic to imagine a crystal rotating at a 
frequency which would be greater, or even close to, the frequency of 
an ultrasonic surface wave, this wide range is chosen to show 
numerically that the wave speed tends to zero with increasing 
$\delta$ without, seemingly, ever reaching it. 
For Silica (SiO$_2$, isotropic), the calculations are conducted 
for a smaller range ($0.0 \le \delta \le 0.1$)
with and without the contribution of the centrifugal acceleration in 
order to show its importance. 

Some theoretical issues are left open such as, the questions of 
existence and uniqueness of a surface wave in a rotating crystal, or 
the question of the general behaviour of the wave speed as a function 
of $\delta$.

\section{Equations of motion}

\subsection{General anisotropy and arbitrary rotation axis}  

Consider a semi-infinite body made of a linearly elastic, anisotropic 
crystal. 
For two-dimensional deformations, the stress-strain relations are 
(Ting 1996, p. 38),
\begin{equation} \label{StressStrain} 
\mbox{\boldmath $\sigma^\text{o}$} 
  = \mathbf{C^o}\mbox{\boldmath $\epsilon^\text{o}$},
\quad \text{or} \quad 
\mbox{\boldmath $\epsilon^\text{o}$}
   = \mathbf{s'} \mbox{\boldmath $\sigma^\text{o}$}; 
\quad 
\mathbf{C^o}\mathbf{s'}
 = \mathbf{I}, 
\end{equation} 
where 
\begin{equation} 
\mbox{\boldmath $\sigma^\text{o}$} = 
 [\sigma_{11}, \sigma_{22}, \sigma_{23}, 
        \sigma_{31}, \sigma_{12}]^\text{T}, \quad 
\mbox{\boldmath $\epsilon^\text{o}$} = 
 [\epsilon_{11}, \epsilon_{22}, 2\epsilon_{23}, 
        2\epsilon_{31}, 2\epsilon_{12}]^\text{T},
\end{equation}
and
\begin{equation} \label{Cs'} 
\mathbf{C^o} = 
 \begin{bmatrix} 
  C_{11} & C_{12} & C_{14} & C_{15} & C_{16} \\ 
         & C_{22} & C_{24} & C_{25} & C_{26} \\ 
         &        & C_{44} & C_{45} & C_{46} \\ 
         &        &        & C_{55} & C_{56} \\ 
         &        &        &        & C_{66} 
 \end{bmatrix} \quad 
\mathbf{s'} = 
 \begin{bmatrix} 
  s'_{11} & s'_{12} & s'_{14} & s'_{15} & s'_{16} \\ 
          & s'_{22} & s'_{24} & s'_{25} & s'_{26} \\ 
          &         & s'_{44} & s'_{45} & s'_{46} \\ 
          &         &         & s'_{55} & s'_{56} \\ 
          &         &         &         & s'_{66} 
 \end{bmatrix}. 
\end{equation} 
Here the $C_{ij}$ are derived, using the Voigt (1910, p.560) 
contracted notation, from the fourth-order elastic stiffness tensor 
and the reduced elastic compliances $s'_{ij}$ are defined by
\eqref{StressStrain}$_3$. 

In the Cartesian orthogonal coordinate system 
$(Ox_1 x_2 x_3) \equiv (O\mathbf{i}\mathbf{j}\mathbf{k})$ 
where $x_2 \ge 0$ is the region 
occupied by the half-space, a surface (Rayleigh) wave traveling 
at speed $v$ and wave number $k$ in the $x_1$-direction with 
attenuation in the $x_2$-direction is described by the following 
mechanical displacement field, 
\begin{equation} \label{u}
\mathbf{u}(x_1,x_2,x_3,t)
   = \mathbf{U}(kx_2) \re^{\ri k(x_1-vt)},
  \quad  \mathbf{U}(\infty) = 0. 
\end{equation} 
From \eqref{StressStrain}$_1$ it follows that the corresponding stress 
components are of similar form. 
In particular, the tractions acting upon the planes $x_2= \const$ can 
be written as
\begin{equation} \label{t}
\sigma_{i2}(x_1,x_2,x_3,t) 
  = \ri k t_i(kx_2) \re^{\ri k(x_1-vt)}. 
\end{equation}

Now consider that the crystal is rotating at a uniform rate $\Omega$
about the direction of a unit vector $\mathbf{w}$.
Then the incremental (time-dependent) equations of motion in the rotating 
frame are (Schoenberg \& Censor 1973)
\begin{equation} \label{motionGeneral}
\text{\textbf{div} } \mbox{\boldmath $\sigma$}
  = \rho \ddot{\mathbf{u}} 
     + 2 \rho \Omega \mathbf{w} \times \dot{\mathbf{u}}
      + \rho \Omega^2 \mathbf{w} \times (\mathbf{w} \times \mathbf{u}),
\end{equation}
where $\rho$ is the mass density and the dot denotes differentiation 
with respect to time.
Note that on the right hand-side of the equations, both the Coriolis 
(second term) and the centripetal (third term) accelerations appear.
Using \eqref{StressStrain}, \eqref{u}, and \eqref{t}, these equations 
can be formulated as a linear homogeneous system of first-order 
differential equations for $\mathbf{U} = [U_1, U_2, U_3]^\text{T}$ and 
$\mathbf{t} = [t_1, t_2, t_3]^\text{T}$:
\begin{equation} \label{motion}
\begin{bmatrix}
  \mathbf{U}' \\
  \mathbf{t}'
\end{bmatrix}
 = \ri \begin{bmatrix}
          \mathbf{N_1} & \mathbf{N_2} \\
          \mathbf{N_3} + \mathbf{\check{J}} + X \mathbf{1}
            &  \mathbf{N_1}^\text{T}  
        \end{bmatrix}
        \begin{bmatrix}
         \mathbf{U} \\
         \mathbf{t}
       \end{bmatrix},
\end{equation}
where $X = \rho v^2$ and the prime denotes differentiation 
with respect to $kx_2$.
Here, $\mathbf{N_1}$, $\mathbf{N_2}  = \mathbf{N_2}^\text{T}$,
and  $\mathbf{N_3}  = \mathbf{N_3}^\text{T}$ are the usual 
submatrices of the fundamental matrix $\mathbf{N}$ 
(Ingebrigsten \& Tonning 1969).
Barnett \& Chadwick (1990) and Ting (1988) give explicit expressions 
of the components of the $\mathbf{N_i}$ in terms of the $C_{ij}$ and 
of the $s'_{ij}$, respectively.
The matrix $\mathbf{\check{J}}$ contains all the information relative 
to the rotation. 
Explicitly (Schoenberg \& Censor 1973), 
\begin{equation}
\check{J}_{ij} 
 = [(\delta_{ij} - w_i w_j)\delta^2 
      - 2 \ri \epsilon_{ijk} w_k \delta]X,
\quad
\delta = \Omega / \omega,
\end{equation}
where $\omega = k v$ is the real frequency of the wave,
$\delta_{ij}$ is the Kronecker operator, and $ \epsilon_{ijk}$ is the
alternator. 
Note that $\mathbf{\check{J}}$ is Hermitian.

Finally, the following boundary conditions apply: vanishing of the 
tractions on the free plane surface $x_2=0$ and of the wave as 
$x_2 \rightarrow \infty$, that is
\begin{equation} \label{BC}
\mathbf{t}(0) = 0, \quad
\mathbf{U}(\infty) = 0, \quad
\mathbf{t}(\infty) = 0.
\end{equation}

\subsection{Orthorhombic crystals} 

Henceforward, we are concerned with orthorhombic crystals whose 
crystallographic axes are aligned with the $Ox_i$.
In that case, we have 
\begin{equation}
\mathbf{N_1} = 
\begin{bmatrix}
                          0           & -1 & 0 \\
             -\frac{s'_{12}}{s'_{11}} &  0 & 0 \\
                          0           &  0 & 0
\end{bmatrix},
\quad
\mathbf{N_2} = 
\begin{bmatrix}
  s'_{66} &  0     &    0     \\
      0   & s'_{22} - \frac{s^{'2}_{12}}{s'_{11}} &    0      \\
      0   &        0                              & s'_{44}
\end{bmatrix}.
\end{equation}
Also, the crystal is assumed to be rotating about a crystallographic 
axis.
Introducing the Hermitian tensor 
$\mathbf{\check{K}^{(1)}}
 =  \mathbf{N_3} + \mathbf{\check{J}} + X \mathbf{1}$
(lower left submatrix in \eqref{motion}), we have
\begin{multline} \label{K(1)X1X2X3} 
\mathbf{\check{K}^{(1)}} =  
\begin{bmatrix}
  X - \frac{1}{s'_{11}} &         0            &         0    \\
        0          & (1+\delta^2)X & -2 \ri \delta X  \\
        0          &  2 \ri \delta X & (1+\delta^2)X-\frac{1}{s'_{55}} 
\end{bmatrix},
\\
\begin{bmatrix}
  (1+\delta^2)X - \frac{1}{s'_{11}} &   0   & 2 \ri \delta X       \\
        0          & X &  0  \\
       -2 \ri \delta X    & 0 & (1+\delta^2)X-\frac{1}{s'_{55}} 
\end{bmatrix},
\\
\begin{bmatrix}
 (1+\delta^2) X - \frac{1}{s'_{11}} &  2 \ri \delta X    &   0  \\
       -2 \ri \delta X      & (1+\delta^2)X  &         0    \\
        0          &  0 &     X-\frac{1}{s'_{55}} 
\end{bmatrix},
\end{multline}
for rotation about the $Ox_1$ axis, $Ox_2$ axis, and $Ox_3$ axis, 
respectively.

\section{Explicit dispersion equations.} 
The Stroh-Barnett-Lothe formalism has beautifully addressed and 
resolved most of the problems arising in linear anisotropic elasticity 
(see the textbook by Ting (1996) for a thorough review.)
Some researchers have however followed other routes for the study of 
SAWs in anisotropic crystals (see Ting (2003$b$) for an exposition of 
several methods.)
For instance, Fu \& Mielke (2002) recently proposed a highly efficient 
numerical scheme to determine the Rayleigh wave speed, which is based 
on a Ricatti equation.
Here the method of the polarization vector 
(Currie 1979; Taziev 1989; Ting 2003$ab$) is presented for rotating 
orthorhombic crystals.
Only simple algebraic manipulations and first integrals (Mozhaev 1995) 
are used, so that the derivation should appeal to the reader who is 
not familiar with the Stroh (1958, 1962) formalism and with the 
orthogonality relations (Barnett \& Lothe 1973).

\subsection{Method of the polarization vector.} 
 
Let us examine the equations of motion \eqref{motion} closer.
They read:
\begin{equation}
- \ri \mathbf{U}' = \mathbf{N_1} \mathbf{U} +  \mathbf{N_2} \mathbf{t},
\quad
- \ri \mathbf{t}' = \mathbf{\check{K}^{(1)}} \mathbf{U}
                          +  \mathbf{N_1}^\text{T} \mathbf{t}.
\end{equation}
Take the scalar product of the first line by $\overline{\mathbf{t}}'$ 
and of the second line by $\overline{\mathbf{U}}'$;
then add, together with the complex conjugate;
finally, integrate between 0 and $\infty$ to obtain:
\begin{equation}
[\overline{\mathbf{U}} \cdot \mathbf{\check{K}^{(1)}} \mathbf{U}
  + \overline{\mathbf{t}} \cdot \mathbf{N_2} \mathbf{t}
   +  \overline{\mathbf{t}} \cdot \mathbf{N_1} \mathbf{U}
    +  \mathbf{t} \cdot \mathbf{N_1} \overline{\mathbf{U}} ]_0^\infty
      =0,
\end{equation}
which, using the boundary conditions \eqref{BC}, reduces to:
$\overline{\mathbf{U}}(0) \cdot \mathbf{\check{K}^{(1)}} \mathbf{U}(0)
   = 0$, a relation first established by Stroh (1958) in the static, 
non-rotating case.

This procedure is easily generalized as follows.
First compute $\mathbf{N}$ (the $6 \times 6$ matrix on the right 
hand-side of the equations of motion \eqref{motion}) to the power $n$, 
where $n$ is any positive or negative integer, as
\begin{equation} 
 \mathbf{N}^n
 =  \begin{bmatrix}
          \mathbf{N_1^{(n)}} & \mathbf{N_2^{(n)}} \\
          \mathbf{\check{K}^{(n)}}
            &  \mathbf{N_1^{(n)}}^\text{T}  
        \end{bmatrix},
\end{equation}
(say).
Of course, the matrices $ \mathbf{N_1^{(n)}}$, $\mathbf{N_2^{(n)}}$, 
and $\mathbf{\check{K}^{(n)}}$ do not coincide with $ \mathbf{N_1}^n$, 
$\mathbf{N_2}^n$, and $\mathbf{\check{K}}^n$, respectively.
The matrix  $\mathbf{N_2^{(n)}}$ is however symmetric just like 
$\mathbf{N_2}$ is, as can be proved by induction (Ting 2003$a$).
Similarly it can  be proved, in essentially the same manner, that 
$\mathbf{\check{K}^{(n)}}$ is Hermitian just like 
$\mathbf{\check{K}^{(1)}}$ is.
Now multiply \eqref{motion} in turn by $\mathbf{N}^{-1}$ and by  
$- \ri \mathbf{N}$ to get
\begin{equation}
\left\{
\begin{array}{ll}
 \ri \mathbf{U} = \mathbf{N_1^{(-1)}} \mathbf{U}'
                        +  \mathbf{N_2^{(-1)}} \mathbf{t}',
\\
 \ri \mathbf{t} = \mathbf{\check{K}^{(-1)}} \mathbf{U}'
                          +  \mathbf{N_1^{(-1)}}^\text{T} \mathbf{t}',
\end{array}
\right.
\quad \text{and} \quad
\left\{
\begin{array}{ll}
 - \ri \mathbf{N_1} \mathbf{U}' - \ri \mathbf{N_2} \mathbf{t}' 
 = \mathbf{N_1^{(2)}} \mathbf{U} +  \mathbf{N_2^{(2)}} \mathbf{t},
\\
- \ri  \mathbf{\check{K}^{(1)}} \mathbf{U}'
  - \ri  \mathbf{N_1}^\text{T}  \mathbf{t}'
   = \mathbf{\check{K}^{(2)}} \mathbf{U}
                          +  \mathbf{N_1^{(2)}}^\text{T} \mathbf{t},
\end{array}
\right.
\end{equation}
Take the scalar product of the first line in the first (second) system 
by $\overline{\mathbf{t}}$ ($\overline{\mathbf{t}}'$) 
and of the second line by $\overline{\mathbf{U}}$ 
($\overline{\mathbf{U}}'$);
then add, together with the complex conjugate;
finally, integrate between 0 and $\infty$ to obtain:
\begin{multline}
[\overline{\mathbf{U}} \cdot \mathbf{\check{K}^{(-1)}} \mathbf{U}
 +\overline{\mathbf{t}} \cdot \mathbf{N_2^{(-1)}} \mathbf{t}
  +\overline{\mathbf{t}} \cdot \mathbf{N_1^{(-1)}} \mathbf{U}
   +\mathbf{t}\cdot \mathbf{N_1^{(-1)}}\overline{\mathbf{U}}]_0^\infty
 =0, \\
[\overline{\mathbf{U}} \cdot \mathbf{\check{K}^{(2)}} \mathbf{U}
 +\overline{\mathbf{t}} \cdot \mathbf{N_2^{(2)}} \mathbf{t}
  +\overline{\mathbf{t}} \cdot \mathbf{N_1^{(2)}} \mathbf{U}
   +\mathbf{t}\cdot \mathbf{N_1^{(2)}}\overline{\mathbf{U}}]_0^\infty
 =0,
\end{multline}
respectively.
Using the boundary conditions \eqref{BC}, we have:
$\overline{\mathbf{U}}(0) \cdot \mathbf{\check{K}^{(-1)}}\mathbf{U}(0) 
  = 0$
and 
$\overline{\mathbf{U}}(0) \cdot \mathbf{\check{K}^{(2)}} \mathbf{U}(0) 
  = 0$.
These steps may be repeated \textit{ad infinitum} for any positive or 
negative power $n$ of $\mathbf{N}$ to give
\begin{equation}  \label{uKu}
\overline{\mathbf{U}}(0) \cdot \mathbf{\check{K}^{(n)}} \mathbf{U}(0) 
  = 0.
\end{equation}
However, because of the Cayley-Hamilton theorem, only five 
$\mathbf{N}^n$ are linearly independent and consequently, \eqref{uKu} 
yields at most five equations.
In a non-rotating frame,  Currie (1979) used the equations written at
$n=1,2,3$ for SAWs in the plane of symmetry of a monoclinic crystal 
and Taziev (1989) used the equations written at $n=1,2,3,4,5$ for 
triclinic crystals.
Recently, Ting (2003$a$) pointed out that the choices $n=-2,-1,1,2,3$ 
lead to simpler expressions.

\subsection{Rotation about $Ox_1$ (propagation direction).} 

When the orthorhombic crystal rotates about the propagation direction 
$Ox_1$, the  matrices $\mathbf{\check{K}^{(-2)}}$ and 
$\mathbf{\check{K}^{(2)}}$ have the following structure,
\begin{equation}
\mathbf{\check{K}^{(n)}} = 
 \begin{bmatrix}
                0     & \check{K}^{(n)}_{12} & \ri K^{(n)}_{13} \\
 \check{K}^{(n)}_{12} &          0           &           0     \\
  - \ri K^{(n)}_{13}  &          0           &           0   
 \end{bmatrix},
\quad
n = -2, 2,
\end{equation}
where $ \check{K}^{(n)}_{12}$ and $K^{(n)}_{13}$ are \textit{real}. 
Explicitly,
\begin{equation}
s'_{11}  \check{K}^{(2)}_{12} = 1 - [s'_{11} + s'_{12}(1+\delta^2)]X, 
\quad
s'_{11} K^{(2)}_{13} = 2 s'_{12} \delta X,
\end{equation}
and 
\begin{align} \label{K(-2)X1}
& [s'_{22} - (s'_{11}s'_{22}-s^{'2}_{12})X] D_1  \check{K}^{(-2)}_{12}
  = 
\nonumber \\ 
& \phantom{12345} - 1 + [s'_{11} - (s'_{12}+s'_{55})(1+\delta^2)]X
+ [s'_{11}(1+\delta^2) + s'_{12}(1-\delta^2)^2]s'_{55}X^2, 
\nonumber  \\
&  s'_{44} D_1 K^{(-2)}_{13} = 2 s'_{55} \delta X,
\\
& D_1 = 1 - (s'_{55}+s'_{66})(1+\delta^2)X
         + (1-\delta^2)^2 s'_{55}s'_{66}X^2.
 \nonumber 
\end{align} 
Equations \eqref{uKu} written at $n=-2,2$, for 
$\mathbf{U}(0)
  = [1, \alpha_1 + \ri \alpha_2, \beta_1 + \ri \beta_2]^\text{T}$ 
give:
\begin{equation}
\check{K}^{(-2)}_{12}  \alpha_1 -  K^{(-2)}_{13} \beta_2 = 0,
\quad
\check{K}^{(2)}_{12}  \alpha_1 -  K^{(2)}_{13} \beta_2 = 0.
\end{equation}
Hence, either \textbf{(a)} 
$\check{K}^{(-2)}_{12} K^{(2)}_{13}
      - \check{K}^{(2)}_{12} K^{(-2)}_{13}=0$ or 
\textbf{(b)} $\alpha_1 =  \beta_2 = 0$.
The first possibility is not a valid secular equation, because it is 
not consistent with the non-rotating case;
in other words, \textbf{(a)} at $\delta = 0$ does not reduce to the 
cubic Rayleigh function for non-rotating orthorhombic crystals,
\begin{equation} \label{cubic}
1 - (2s'_{11} + s'_{66})X + s'_{11}(s'_{11} - s'_{22} + 2s'_{66})X^2
  + s'_{11}(s'_{11}s'_{22} - s^{'2}_{12} - s'_{11}s'_{66})X^3 = 0.
\end{equation} 
Thus \textbf{(b)} applies and 
\begin{equation} \label{polarizationX1}
\mathbf{U}(0) = [1,  \ri \alpha_2,  \beta_1]^\text{T}.
\end{equation} 
The directions of the vectors $\mathbf{i} + \beta_1 \mathbf{k}$ and 
$\alpha_2 \mathbf{j}$ are conjugate directions with respect to the 
polarization ellipse of the wave at the free surface 
(Boulanger \& Hayes 1993). 
In fact, because they are orthogonal, they are along the principal 
axes of the ellipse.
Now we compute $\alpha_2$ and $\beta_1$.

The  matrices $\mathbf{\check{K}^{(-1)}}$, $\mathbf{\check{K}^{(1)}}$, 
and $\mathbf{\check{K}^{(3)}}$ have the following structure,
\begin{equation}
\mathbf{\check{K}^{(n)}} = 
 \begin{bmatrix}
 \check{K}^{(n)}_{11} & 0 &      0                   \\
 0 & \check{K}^{(n)}_{22} &  \ri K^{(n)}_{23}        \\
 0 &    - \ri K^{(n)}_{23}    &  \check{K}^{(n)}_{33}
 \end{bmatrix},
\quad
n = -1, 1, 3,
\end{equation}
where $ \check{K}^{(n)}_{11}$, $ \check{K}^{(n)}_{22}$, 
$\check{K}^{(n)}_{33}$, and $K^{(n)}_{23}$ are \textit{real}. 
Explicitly,
\begin{align}
& D_1  \check{K}^{(-1)}_{11}  = 
 - [1 + \delta^2 - s'_{55}(1-\delta^2)^2  X]X, 
&&  s'_{44} \check{K}^{(-1)}_{33} = 1,
\nonumber  \\
& [s'_{22} - (s'_{11}s'_{22}-s^{'2}_{12})X]  \check{K}^{(-1)}_{22}
    = 1 - s'_{11} X,
&&K^{(-1)}_{23} = 0,
\end{align} 
(where $D_1$ is defined in \eqref{K(-2)X1}$_3$);
the components of $\mathbf{\check{K}^{(1)}}$ are given in 
\eqref{K(1)X1X2X3}$_1$; 
and 
\begin{align}
& s'_{11} \check{K}^{(3)}_{11}  = 
  \frac{s'_{66}}{s'_{11}} - 2\frac{s'_{12}}{s'_{11}}
    + [2(s'_{12} - s'_{66}) + \frac{s^{'2}_{12}}{s'_{11}}(1+\delta^2)]X
      + s'_{11}s'_{66}X^2, 
\nonumber  \\
&  s'_{11} \check{K}^{(3)}_{22}  = 
  -1 + [s'_{11} + 2s'_{12}(1+\delta^2)]X 
    + [(s'_{11} s'_{22} - s^{'2}_{12})(1+\delta^2)^2
 + 4s'_{11}s'_{44}\delta^2]X^2,
\nonumber  \\
& s'_{55} \check{K}^{(3)}_{33}  = 
  \frac{s'_{44}}{s'_{55}} - 2 s'_{44}(1+\delta^2)X
    + s'_{55}[s'_{44}(1+\delta^2)^2 + 4 (s'_{22}
         - \frac{s^{'2}_{12}}{s'_{11}})\delta^2]X^2, 
\\
& s'_{11} K^{(3)}_{23}  = 
  - 2\delta X [s'_{12} -  s'_{11} \frac{s'_{44}}{s'_{55}} 
       +  (s'_{11}s'_{22}-s^{'2}_{12} + s'_{11}s'_{44})(1+\delta^2)X].
\nonumber 
\end{align} 
Substitution of \eqref{polarizationX1} into equations \eqref{uKu} 
written at $n=-1,1,3$ gives:
\begin{equation}
\check{K}^{(n)}_{11}
 +  \check{K}^{(n)}_{33} \beta_1^2
  + \check{K}^{(n)}_{22}  \alpha_2^2
   + 2 K^{(n)}_{23}  \alpha_2 \beta_1 = 0,
\quad n=-1,1,3.
\end{equation}
These three equations are rewritten as $F_{ik} g_k = -h_i$ with
\begin{equation}
\mathbf{F} = 
\begin{bmatrix}
 \check{K}^{(-1)}_{33} & \check{K}^{(-1)}_{22} & K^{(-1)}_{23}  \\ 
  \check{K}^{(1)}_{33} &  \check{K}^{(1)}_{22} &  K^{(1)}_{23}  \\ 
  \check{K}^{(3)}_{33} &  \check{K}^{(3)}_{22} &  K^{(3)}_{23}  
  \end{bmatrix},
\quad
\mathbf{g} = 
\begin{bmatrix}
   \beta_1^2  \\ 
  \alpha_2^2  \\ 
  2 \alpha_2 \beta_1 
  \end{bmatrix},
\quad
\mathbf{h} = 
\begin{bmatrix}
 \check{K}^{(-1)}_{11}  \\ 
  \check{K}^{(1)}_{11}  \\ 
  \check{K}^{(3)}_{11}  
  \end{bmatrix}.
\end{equation}
Let $\Delta = \text{det } \mathbf{F}$, and $\Delta_k$ be the 
determinant of the matrix obtained from $\mathbf{F}$ by replacing the 
$k$-th column with $\mathbf{h}$.
Then the solution to $\mathbf{Fg} = \mathbf{h}$ is 
$g_k = - \Delta_k / \Delta$.
But $g_3^2 = 4 g_1 g_2$, that is
\begin{equation} \label{secularX1}
\Delta_3^2 - 4 \Delta_1 \Delta_2 = 0,
\end{equation}
the \textit{explicit dispersion equation for Rayleigh waves in rhombic 
crystals rotating about $Ox_1$}.
It is a polynomial of degree 12 in $X = \rho v^2$ and of degree 10 in 
$\delta^2$. 
At $\delta = 0$, $K^{(n)}_{23}=0$, so that $\Delta_1 = \Delta_2 =0$, 
while $\Delta_3$ factorizes into the product of a quadratic in $X$ and 
the cubic \eqref{cubic}.

\subsection{Rotation about $Ox_2$ (normal to the free plane).} 
 
When the orthorhombic crystal rotates about  $Ox_2$, the normal to the 
free plane, the  matrices $\mathbf{\check{K}^{(-2)}}$ and 
$\mathbf{\check{K}^{(2)}}$ have the following structure,
\begin{equation}
\mathbf{\check{K}^{(n)}} = 
 \begin{bmatrix}
                0     & \check{K}^{(n)}_{12} &    0           \\
 \check{K}^{(n)}_{12} &         0            &  \ri K^{(n)}_{23} \\
               0      &   - \ri K^{(n)}_{23} &           0   
 \end{bmatrix},
\quad
n = -2, 2,
\end{equation}
where $ \check{K}^{(n)}_{12}$ and $K^{(n)}_{23}$ are \textit{real}. 
Explicitly,
\begin{equation}
s'_{11}  \check{K}^{(2)}_{12} = 1 - [s'_{12} + s'_{11}(1+\delta^2)]X, 
\quad
K^{(2)}_{23} = -2 \delta X,
\end{equation}
and 
\begin{align} \label{K(-2)X2}
& s'_{11} (1 - s'_{66}X) D_2  \check{K}^{(-2)}_{12}  = 
\nonumber \\ 
& \phantom{12345} - 1 + [s'_{12} + (s'_{11}+s'_{55})(1+\delta^2)]X
           - [s'_{12}(1+\delta^2) + s'_{11}(1-\delta^2)^2]s'_{55}X^2, 
\nonumber  \\
&  \frac{s'_{44}}{s'_{55}} s'_{11} D_2 K^{(-2)}_{23}
          = -2 s'_{12} \delta X,
\\
& s'_{11}D_2 
 = s'_{22} - [s'_{22}(s'_{11}+s'_{55})  - s^{'2}_{12}](1+\delta^2)X 
            + [s'_{12}(1+\delta^2) + s'_{11}(1-\delta^2)^2] s'_{55}X^2.
 \nonumber 
\end{align} 
Equations \eqref{uKu} written at $n=-2,2$, for 
$\mathbf{U}(0)
 = [1, \alpha_1 + \ri \alpha_2, \beta_1 + \ri \beta_2]^\text{T}$ 
give:
\begin{equation}
\check{K}^{(-2)}_{12}  \alpha_1
   + K^{(-2)}_{23} (\alpha_2 \beta_1- \alpha_1 \beta_2) = 0,
\quad
\check{K}^{(2)}_{12}  \alpha_1
   + K^{(2)}_{23} (\alpha_2 \beta_1- \alpha_1 \beta_2) = 0.
\end{equation}
Hence, either \textbf{(a)} 
$\check{K}^{(-2)}_{12} K^{(2)}_{23}
  - \check{K}^{(2)}_{12} K^{(-2)}_{23}=0$, 
or \textbf{(b)} $\alpha_1 =  \alpha_2 = 0$, or 
\textbf{(c)} $\alpha_1 =  \beta_1 = 0$.
The first possibility is not a valid secular equation, because it is 
not consistent with the non-rotating case (that is \textbf{(a)} does 
not reduce to the cubic \eqref{cubic} at $\delta = 0$.)
It can be checked that in Case \textbf{(b)}, the equations \eqref{uKu} 
written at $n=-1,1,3,$ lead to a secular equation which vanishes 
identically at $\delta=0$, so that \textbf{(b)} is also dismissed.
Thus \textbf{(c)} remains, and 
\begin{equation} \label{polarizationX2}
\mathbf{U}(0) = [1,  \ri \alpha_2,  \ri \beta_2]^\text{T}.
\end{equation} 
The directions of the vectors $\alpha_2 \mathbf{j}+\beta_2 \mathbf{k}$ 
and $\mathbf{i}$ are along the principal axes of the polarization 
ellipse at the free surface.

The  matrices $\mathbf{\check{K}^{(-1)}}$, $\mathbf{\check{K}^{(1)}}$, 
and $\mathbf{\check{K}^{(3)}}$ have the following structure,
\begin{equation}
\mathbf{\check{K}^{(n)}} = 
 \begin{bmatrix}
 \check{K}^{(n)}_{11} & 0 & \ri K^{(n)}_{13}       \\
 0 & \check{K}^{(n)}_{22} &   \\
 - \ri K^{(n)}_{13}       &    0  &  \check{K}^{(n)}_{33}
 \end{bmatrix},
\quad
n = -1, 1, 3,
\end{equation}
where $ \check{K}^{(n)}_{11}$, $ \check{K}^{(n)}_{22}$, 
$\check{K}^{(n)}_{33}$, and $K^{(n)}_{13}$ are \textit{real}. 
Explicitly,
\begin{align}
& (1 - s'_{66})X  \check{K}^{(-1)}_{11}  = - X,  
\quad s'_{44} \check{K}^{(-1)}_{33} = 1, 
\quad K^{(-1)}_{13} = 0, 
\nonumber  \\
& s'_{11} D_2 \check{K}^{(-1)}_{22}
 = 1 - (s'_{11} + s'_{55})(1+\delta^2) 
                           + s'_{11} s'_{55}(1-\delta^2)^2  X^2,
\end{align} 
(where $D_2$ is defined in \eqref{K(-2)X2}$_3$);
the components of $\mathbf{\check{K}^{(1)}}$ are given in 
\eqref{K(1)X1X2X3}$_2$; 
and 
\begin{align}
& s'_{11} \check{K}^{(3)}_{11}  = 
  \frac{s'_{66}}{s'_{11}} - 2\frac{s'_{12}}{s'_{11}}
  + [\frac{s^{'2}_{12}}{s'_{11}}+2(s'_{12} - s'_{66})(1+\delta^2)]X 
\nonumber \\
& \phantom{123456789012345678901234567890}
        + s'_{11}[s'_{66}(1+\delta^2)^2 + 4s'_{44}\delta^2]X^2, 
\nonumber  \\
&  s'_{11} \check{K}^{(3)}_{22}  = 
  -1 + [2s'_{12} + s'_{11}(1+\delta^2)]X 
    + (s'_{11} s'_{22} - s^{'2}_{12})X^2,
\\
& s'_{55} \check{K}^{(3)}_{33}  = 
  \frac{s'_{44}}{s'_{55}} - 2 s'_{44}(1+\delta^2)X
    + s'_{55}[s'_{44}(1+\delta^2)^2 + 4 s'_{66}\delta^2]X^2, 
\nonumber  \\
& s'_{11} K^{(3)}_{23}  = 
   2\delta X [s'_{12} - s'_{66} - s'_{11} \frac{s'_{44}}{s'_{55}} 
       +  s'_{11}(s'_{44} + s'_{66})(1+\delta^2)X].
\nonumber 
\end{align} 
Substitution of \eqref{polarizationX2} into equations \eqref{uKu} 
written at $n=-1,1,3$ gives:
\begin{equation}
\check{K}^{(n)}_{11}
 +  \check{K}^{(n)}_{33} \beta_2^2
  + \check{K}^{(n)}_{22}  \alpha_2^2
   - 2 K^{(n)}_{13}  \beta_2 = 0,
\quad n=-1,1,3.
\end{equation}
These three equations are rewritten as $F_{ik} g_k = -h_i$ with
\begin{equation}
\mathbf{F} = 
\begin{bmatrix}
 \check{K}^{(-1)}_{33} & \check{K}^{(-1)}_{22} & K^{(-1)}_{13}  \\ 
  \check{K}^{(1)}_{33} &  \check{K}^{(1)}_{22} &  K^{(1)}_{13}  \\ 
  \check{K}^{(3)}_{33} &  \check{K}^{(3)}_{22} &  K^{(3)}_{13}  
  \end{bmatrix},
\quad
\mathbf{g} = 
\begin{bmatrix}
   \beta_2^2  \\ 
  \alpha_2^2  \\ 
  -2 \beta_2 
  \end{bmatrix},
\quad
\mathbf{h} = 
\begin{bmatrix}
 \check{K}^{(-1)}_{11}  \\ 
  \check{K}^{(1)}_{11}  \\ 
  \check{K}^{(3)}_{11}  
  \end{bmatrix}.
\end{equation}
Let $\Delta = \text{det } \mathbf{F}$, and $\Delta_k$ be the 
determinant of the matrix obtained from $\mathbf{F}$ by replacing the 
$k$-th column with $\mathbf{h}$.
Then the solution to $\mathbf{Fg} = \mathbf{h}$ is 
$g_k = - \Delta_k / \Delta$.
But $g_3^2 = 4 g_1$, that is
\begin{equation} \label{secularX2}
\Delta_3^2 + 4 \Delta_1 \Delta = 0,
\end{equation}
the \textit{explicit dispersion equation for Rayleigh waves in rhombic 
crystals rotating about $Ox_2$}.
It is a polynomial of degree 12 in $X = \rho v^2$ and of degree 10 in 
$\delta^2$. 
At $\delta = 0$, $K^{(n)}_{13}=0$, so that $\Delta = \Delta_1 =0$, 
while $\Delta_3$ factorizes into the product of a quadratic in $X$
and the cubic \eqref{cubic}.

\begin{figure}
\begin{centering}
\epsfig{figure=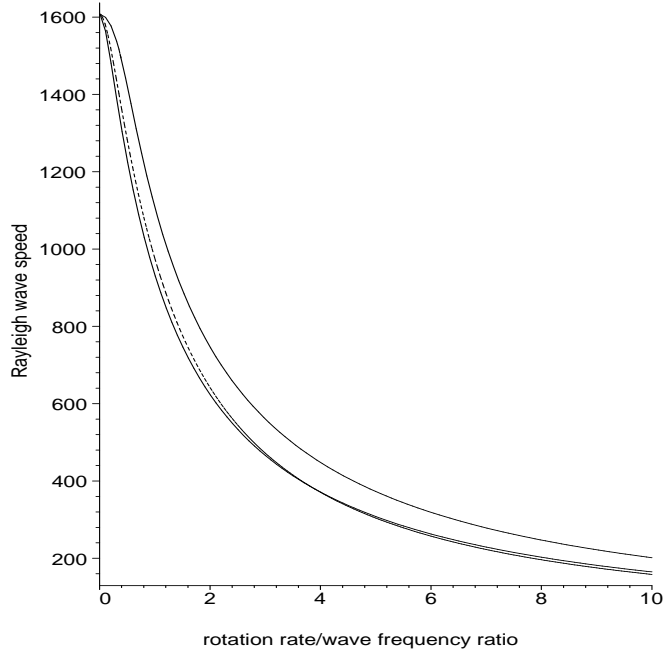, height=8.5cm, width=8.5cm}
\caption{Rayleigh wave speeds for $\alpha$-Iodic Acid (rhombic) 
rotating about 
$Ox_1$ (bottom solid curve), $Ox_2$ (top solid curve), 
or $Ox_3$ (dotted curve).}
\end{centering}
\end{figure}

Figure 1 shows the influence of $\delta$ upon the surface wave speed 
for  $\alpha$-Iodic Acid (HIO$_3$, rhombic).
The elastic stiffnesses ($10^{10}$ N/m$^2$) are 
(Royer \& Dieulesaint 1996): $C_{11} = 3.01$, $C_{12} = 1.61$, 
$C_{22} = 5.80$, $C_{44} = 1.69$,  $C_{55} = 2.06$, and 
$C_{66} = 1.58$;
the mass density is 4640 kg/m$^3$.
We see that the Rayleigh wave speed is a monotone decreasing function 
of $\delta$.
When the crystal rotates about the $x_2$-axis (solid top curve), the 
speed is greater at a given $\delta$ than when it rotates about the 
$x_1$-axis (solid bottom curve) or the $x_3$-axis (dotted curve).

\subsection{Rotation about $Ox_3$.} 
 
The case where the orthorhombic crystal rotates about an axis 
orthogonal to the direction of propagation and to the normal to the 
free surface has been treated elsewhere (Destrade 2003) as
a special case of a monoclinic crystal.
It turns out that the in-plane strain decouples from anti-plane 
strain, and that the displacement is of the form,
\begin{equation} \label{polarizationX3}
\mathbf{U}(0) = [1,  \ri \alpha_2, 0]^\text{T},
\end{equation} 
where $\alpha_2$ is real.
Explicitly, the wave propagates at a speed which is a root of the 
following dispersion equation:
\begin{equation}
(h_1F_{23} - h_2 F_{13})^2
 - (F_{12}h_2 - F_{22}h_1)(F_{12}F_{23} - F_{22}F_{13})=0,
\end{equation}
with 
\begin{equation}
\begin{array}{ll}
 F_{12} = 2s'_{12}\delta X, \quad
   F_{13} = 1 - (s'_{11}+s'_{66})(1+\delta^2)X 
            + s'_{11}s'_{66}(1-\delta^2)^2 X^2, 
\\
F_{22} = -2s'_{11}\delta X, \quad
   F_{23} =  s'_{11}(1+\delta^2)X, 
\\
h_1 = [s'_{22}(1+\delta^2) 
       - (s'_{11}s'_{22}-s^{'2}_{12})(1-\delta^2)^2X]X, \quad
h_2 = 1 - s'_{11}(1+\delta^2)X,
\end{array} 
\end{equation}
and the mechanical displacement at the free surface is given by 
\eqref{polarizationX3}, with 
$\alpha_2^2 = (F_{12}h_2 - F_{22}h_1)/(F_{12}F_{23} - F_{22}F_{13})$.

The dispersion equation is a polynomial of degree 6 in $X=\rho v^2$ 
and in $\delta^2$.  
In the \textit{isotropic} case, it reduces to $a_i \xi^i =0$ 
($i=0,1,\ldots,6$), where $\xi = \rho v^2 / C_{66}$, 
$\alpha = C_{66} / C_{11}$, and
\begin{align}
& a_6 = 
 (1-\delta^2)^4
   [(1+\alpha)^2(1+\delta^2)^2 - 4\alpha(1-\delta^2)^2], 
\nonumber \\
& a_5 = 
 -16 (1-\alpha)(1+\delta^2)(1-\delta^2)^2
   [(1-\alpha)(1+\delta^4) + 2(1+3\alpha)\delta^2], 
\nonumber \\
& a_4 = 
 16 (1-\alpha)^2(1+\delta^2)^2
   [(7-2\alpha)(1+\delta^4) + 2(1+2\alpha)\delta^2], 
\nonumber \\
& a_3 = 
 -32 (1-\alpha)^2(1+\delta^2)
   [(13-9\alpha)(1+\delta^4) + 2(7-11\alpha)\delta^2], 
\\
& a_2 = 
 64 (1-\alpha)^2
   [(13-16\alpha+4\alpha^2)(1+\delta^2)^2 - 4\delta^2], 
\nonumber \\
& a_1 = 
 -256 (3-2\alpha)(1-\alpha)^3(1+\delta^2), 
\nonumber \\
& a_0 = 
 256 (1-\alpha)^4.
\nonumber
\end{align}
This equation corresponds to the rationalized form of the dispersion 
equation obtained by Clarke \& Burdess (1994). 
In the non-rotating case, $\delta=0$ and the sextic secular 
equation reduces to the (squared) cubic of Rayleigh (1885):
$\xi^3 - 8\xi^2 + (24-16\alpha)\xi - 16(1-\alpha)=0$. 
Finally in the \textit{incompressible isotropic} case, $\alpha = 0$ 
and the sextic equation factorizes into the product of the two 
following cubics,
\begin{equation}
(1+\delta^2) (1 \pm \delta)^4 \xi^3 
  - 8(1+\delta^2)(1 \pm \delta)^2 \xi^2 
+ 8(3 \pm 2\delta + 3 \delta^2) \xi  - 16 =0.
\end{equation}

\begin{figure}
\begin{centering}
\epsfig{figure=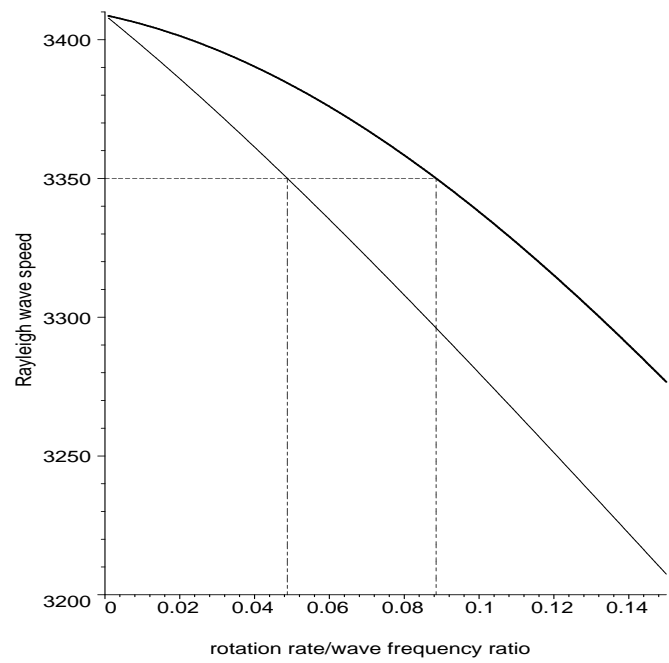, height=8.5cm, width=8.5cm}
\caption{Rayleigh wave speeds for Silica (isotropic) 
rotating about $Ox_3$ calculated with (thick curve) and without 
(thin curve) the contribution of the centrifugal force.}
\end{centering}
\end{figure}

Figure 2 shows the influence of $\delta$ upon the wave speed (thick 
curve) for Silica (SiO$_2$, isotropic) rotating about the normal to
the directions of propagation and of attenuation.
The elastic stiffnesses ($10^{10}$ N/m$^2$) are 
(Royer \& Dieulesaint 1996): $C_{11} = 7.85$ and $C_{12} = 1.61$;
the mass density is 2203 kg/m$^3$.
The thin curve corresponds to the dispersion curve obtained when the 
centrifugal acceleration is omitted in the equations of motion 
\eqref{motionGeneral}.
We see that even for small $\delta$, this term plays an important role:
for instance, the speed is reduced from 3409 m/s in the non-rotating
case to 3350 m/s (horizontal dotted line) for $\delta = 0.08852$ with 
the full equations of motion, and for $\delta = 0.04878$ without the 
centripetal acceleration, a relative difference of 45\% for the 
expected value of $\delta$!
Similarly, Grigor'evski\u{i} \textit{et al.} (2000) argued that one of 
the bulk waves disappeared in a rotating isotropic media at: 
$\delta = \textstyle{\frac{1}{2}}$; 
in fact, as shown by Schoenberg \& Censor (1973), this phenomenon 
occurs at `resonance': $\delta = \Omega/\omega = 1$.

\subsection{Remark: rotation about any direction in $x_3=0$ plane.}

Note that the method of the polarization vector may also be carried 
out fully in the case where the crystal rotates about any axis in the
($x_1x_2$) plane.
Then the secular equation is obtained as a nonlinear combination of 
determinants of $5 \times 5$ matrices, as follows.

When the unit vector $\mathbf{w}$ along the axis of rotation is such 
that $w_3=0$ and $w_1 w_2 \ne 0$, then the matrices 
$\mathbf{\check{K}^{(n)}}$ have the following structure,
\begin{equation}
\mathbf{\check{K}^{(n)}} = 
 \begin{bmatrix}
 \check{K}^{(n)}_{11} & \check{K}^{(n)}_{12} & \ri K^{(n)}_{13} \\
 \check{K}^{(n)}_{12} & \check{K}^{(n)}_{22} & \ri K^{(n)}_{23} \\
  -\ri K^{(n)}_{13}   &   - \ri K^{(n)}_{23} & \check{K}^{(n)}_{33}    
 \end{bmatrix},
\end{equation}
where $\check{K}^{(n)}_{11}$, $\check{K}^{(n)}_{22}$, 
$\check{K}^{(n)}_{33}$, $\check{K}^{(n)}_{12}$, $K^{(n)}_{13}$, 
and $K^{(n)}_{23}$ are \textit{real}.

Substitution of $\mathbf{U}(0)
  = [1, \alpha_1 + \ri \alpha_2, \beta_1 + \ri \beta_2]^\text{T}$ 
into equations \eqref{uKu} written at five different $n$ gives:
\begin{align}
& \check{K}^{(n)}_{22}g_1
 + \check{K}^{(n)}_{33}g_2
  + \check{K}^{(n)}_{12}g_3 
   + K^{(n)}_{13}g_4 
    + K^{(n)}_{23}g_5 
     = - \check{K}^{(n)}_{11}, 
\nonumber \\
& g_1 = \alpha_1^2 + \alpha_2^2, \: 
   g_2 =  \beta_1^2 +  \beta_2^2,  \:
    g_3 = 2\alpha_1, \:
     g_4 =-2\beta_2^2, \:
      g_5 = 2(\alpha_2 \beta_1 - \alpha_1 \beta_2).
\end{align}
The nonhomogeneous linear system of five equations is solved for the 
five unknowns $g_i$. 
But these quantities are related through
\begin{equation} 
g_1^2 g_4 + g_3^2 g_2 + g_5^2 - g_1 g_3 g_5 - 4g_2 g_4 = 0,
\end{equation}
the \textit{dispersion equation for Rayleigh waves in rhombic 
crystals rotating about any axis in the ($x_1x_2$) plane}.

Explicitly, the components of the $\mathbf{\check{K}^{(n)}}$ 
matrices are however too long to reproduce here.
Also, the corresponding set-ups are unlikely to be of experimental 
relevance.



\end{document}